\documentclass[conference,letterpaper]{IEEEtran}
\usepackage{setspace}
\usepackage{microtype}
\usepackage{type1cm}
\usepackage{url}
\usepackage{amsmath}
\usepackage{amsthm}
\usepackage{amssymb}
\usepackage{mathtools}
\usepackage{bm}
\usepackage{psfrag}
\usepackage{graphicx}
\usepackage{xspace}
\usepackage{xcolor}
\usepackage{float}
\usepackage[caption=false,font=footnotesize]{subfig}
\usepackage{cite}
\usepackage{floatflt}
\IEEEoverridecommandlockouts

\newtheorem{proposition}{Proposition}
\DeclareMathOperator*{\argmin}{arg\,min}
\DeclareMathOperator{\expop}{E}

\newcommand{\thrp}{\mathsf{S}}
\newcommand{\pwr}{\mathcal{E}}
\newcommand{\btcs}{\mathsf{B}}

\begin{document}
\title{On the Throughput/Bit-Cost Tradeoff\\in CSMA Based Cooperative Networks}

\author{
\IEEEauthorblockN{Georg B\"ocherer\IEEEauthorrefmark{1}
and Rudolf Mathar\IEEEauthorrefmark{1}
}
\\
\IEEEauthorblockA{\IEEEauthorrefmark{1}Institute for Theoretical Information
Technology\\
RWTH Aachen University,
52056 Aachen, Germany\\ Email: \{boecherer,mathar\}@ti.rwth-aachen.de}
\thanks{This work has been supported by the UMIC Research Centre, RWTH
Aachen University.}
}
\maketitle

\begin{abstract}
Wireless local area networks (WLAN) still suffer from a severe performance discrepancy between different users in the uplink. This is because of the spatially varying channel conditions provided by the wireless medium. Cooperative medium access control (MAC) protocols as for example CoopMAC were proposed to mitigate this problem. In this work, it is shown that cooperation implies for cooperating nodes a tradeoff between throughput and bit-cost, which is the energy needed to transmit one bit. The tradeoff depends on the degree of cooperation. For carrier sense multiple access (CSMA) based networks, the throughput/bit-cost tradeoff curve is theoretically derived. A new distributed CSMA protocol called fairMAC is proposed and it is theoretically shown that fairMAC can asymptotically achieve any operating point on the tradeoff curve when the packet lengths go to infinity. The theoretical results are validated through Monte Carlo simulations.
\end{abstract}

\section{Introduction}
The motivation for this work is the performance discrepancy for different users in WLAN uplinks as observed in \cite{Heusse2003}.
Cooperation in wireless networks has drawn a lot of
attention in order to mitigate throughput discrepancy between users in wireless
networks. Based on the early results presented in \cite{Cover1979}, the authors in
\cite{Sendonaris2003,Sendonaris2003a} illustrate that cooperation
between two co-located users can be beneficial for both users when transmitting over fading channels. Several works propose distributed
protocols to coordinate cooperation at the MAC layer, for instance
\textit{r}DCF~\cite{Zhu2006a} and CoopMAC~\cite{Liu2007}. Both protocols enable
two-hop transmission as an alternative to direct transmission for WLAN. These
protocols also coordinate cooperation on the physical layer~\cite{Liu2008}.
The benefits of cooperation for the whole network have been discussed
in~\cite{Korakis2007a,Narayanan2007}. In~\cite{Zhu2006a,Bletsas2006,Liu2007}, the authors propose to select the best
relay for each transmission separately. However, if one node is determined as
the best relay for many nodes, its energy consumption will be very high compared
to other nodes. In~\cite{Bocherer2008} we investigate distributed cooperative
protocols for two users based on CSMA, where both users were constrained to achieve
same throughput with same energy consumption, i.e., full fairness. This was
achieved by individual transmission power adaption for each user. However the extension to scenarios
with many users is rather unrealistic since it would require centralized power allocation, which is difficult to implement in ad-hoc networks.

In this work, we choose a different approach and restrict all transmitters to the same average transmit power during transmission. We impose throughput fairness as in \cite{Bocherer2008}, i.e., on the long term, each node effectively transmits information at the same rate to the common access point (AP). For evaluation, we consider the effective throughput \textit{and} the resulting bit-cost in terms of average energy per transmitted data. 
\begin{itemize}
\item We identify a throughput/bit-cost tradeoff in cooperative networks: a potential helper increases his own \emph{throughput} by cooperating, but he also increases his \emph{bit-cost}.
\item We analytically derive formulas for the throughput/bit-cost tradeoff curve that results from timesharing between CSMA based CoopMAC~\cite{Liu2007} and conventional CSMA based Direct Link, where all nodes transmit directly to the AP.
\item We propose a distributed protocol called fairMAC and show, both theoretically and by Monte Carlo simulation, that fairMAC can asymptotically reach the tradeoff curve when the packet lengths go to infinity.
\end{itemize}

The remainder of this paper is organized as follows. In Section~\ref{sec:model}, we provide our system model. In Section~\ref{sec:keyidea}, we introduce the main topics of our work in a simplified setup. We define the new protocol fairMAC in Section~\ref{sec:protocol} and we analyze it theoretically in Section~\ref{sec:analysis}. Finally, we validate our theoretical results through simulation in Section~\ref{sec:simulation}.

\section{System Model}
\label{sec:model}
We consider a network of $N$ nodes that seek to transmit their data to a common
AP. For each pair of nodes $k,l$ of the network, we associate with the transmission from $k$ to $l$ the achievable rate $R_{kl}$ in bit/s. We denote by $R_k$ the achievable rate for direct transmission from node $k$ to the AP. We normalize the amount of data per packet to $1$~bit. The packet length for a transmission from node $k$ to the AP is then given by $1/R_{k}$. 
For the cooperative protocols CoopMAC and fairMAC (to be
introduced in this work), some nodes have the possibility to transmit their packets to the AP via a helper. Following \cite{Liu2007}, the helper is chosen such that the overall transmission length is minimized: $h$ can help $k$ if and only if
\begin{align}
h=\argmin_{l\in [1,N]}\frac{1}{R_{kl}}+\frac{1}{R_l}\quad
\text{and}\quad
\frac{1}{R_{kh}}+\frac{1}{R_h}<\frac{1}{R_k}\label{eq:helper}.
\end{align}
If such an $h$ exists for node $k$, we denote it by $h_k$. It is the best relay of $k$ for two-hop transmission and transmitting from $k$ via $h_k$ to the AP takes less time than transmitting directly from $k$ to the AP. We assume that node~$k$ knows the rate~$R_k$ and if it has a helper $h_k$ according
to \eqref{eq:helper}, it also knows $R_{kh_k}$. We have a quasi-static environment in mind where a part of the nodes continuously experiences a
channel much worse than other nodes. We therefore assume that the rates of the links remain constant over the period of interest. 

The communications setup is throughout the paper as follows: each transmitted packet contains 1~bit of information. All nodes have an infinite amount of data that they want to transmit to a common AP. All nodes are restricted to the same transmit power $\pwr$ during transmission. The investigated strategies aim to guarantee the same effective throughput on the long term to all nodes independent of their transmission rates.

\section{Throughput/Bit-Cost Tradeoff}
\label{sec:keyidea}
In this introductory section, we give a short overview over the main topics of this work: we first define throughput and bit-cost. We then present the tradeoff between throughput and bit-cost in cooperative networks.
\subsection{Throughput and Bit-Cost}
\label{subsec:throughputBitCost}
The \emph{throughput} $\thrp_k$ of node $k$ is the average amount of data bits per time that node $k$ successfully transmits. Only data belonging to $k$ is taken into account; data that $k$ forwards for other nodes does not contribute to the throughput $\thrp_k$. Let $\bar{\pwr}_k$ denote the \emph{average} power of node $k$ ($\bar{\pwr}$ is given by $\mathrm{transmit\,power\,} \pwr\times \mathrm{transmission\,time}/\mathrm{overall\,time}$). In contrast to the throughput $\thrp_k$, power spent while forwarding data of other nodes \emph{does} contribute to $\bar{\pwr}_k$. We define the \emph{bit-cost} $\btcs_k$ of $k$ as
\begin{align}
\btcs_k=\frac{\bar{\pwr}_k}{\thrp_k}
\end{align}
i.e., it measures the average amount of energy that node $k$ has to spend to successfully transmit one own data bit.

For exposition and comparison, we consider in this section \emph{Round Robin} as a centralized time division multiple access (TDMA) strategy. In a network of $N$ nodes scheduled with Round Robin, the nodes transmit one after each other in a circular order. Denote by $s_k$ the \emph{travel time} of one bit of node $k$ and denote by $t_k$ the \emph{transmission time} of node $k$, i.e., the overall time that node $k$ is transmitting in one round. If node $k$ is transmitting directly to the AP and does not forward data of other nodes, then $s_k=t_k=1/R_k$. If node $k$ is transmitting directly to the AP and is forwarding data of the number of $H_k$ other nodes per round, then $s_k=1/R_k$ and $t_k=(H_k+1)/R_k$. If node $k$ transmits via node $h_k$, then $s_k=1/R_{kh_k}+1/R_{h_k}$ and $t_k=1/R_{kh_k}$. Throughput and bit-cost of node $k$ are thus given by
\begin{align}
\thrp_k = \frac{1 \mathrm{bit}}{\sum_{k=1}^N s_k}
\quad
\text{and}
\quad
\btcs_k = \frac{\bar{\pwr}_k}{\thrp_k}=\frac{\frac{t_k \pwr}{\sum_{k=1}^N s_k}}{\frac{1 \mathrm{bit}}{\sum_{k=1}^N s_k}}=\frac{t_k \pwr}{1 \mathrm{bit}}\label{eq:RRperformance}
\end{align}
Note that $\thrp_k=\thrp_l$ for all $k,l=1,\dotsc,N$. We can thus omit the index and simply refer to throughput $\thrp$, but we have to keep in mind that $\thrp$ is the throughput per node and not the throughput sum over all nodes in the network.

\subsection{A Toy Example}
\begin{figure}
\centering
\footnotesize
\psfrag{node1}{$n_1$}
\psfrag{node2}{$n_2$}
\psfrag{node3}{$n_3$}
\psfrag{node4}{AP}
\psfrag{1}{$1$}
\psfrag{3}{$3$}
\includegraphics[width=0.3\textwidth]{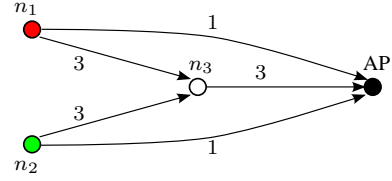}
\caption{A simple network with $3$ nodes and one AP. According to \eqref{eq:helper}, node $n_3$ is a potential helper for both node $n_1$ and node $n_2$.}
\label{fig:toyExample}
\end{figure}
\begin{figure}
\centering
\footnotesize
\psfrag{alpha_coop}{}
\psfrag{alpha_direct}{}
\psfrag{alpha}{$\alpha=5/12$}
\psfrag{Q = 0}{$\alpha=0$}
\psfrag{Q = 1}{$\alpha=5/12$}
\psfrag{Q = 2}{$\alpha=1$}
\includegraphics[width=0.4\textwidth]{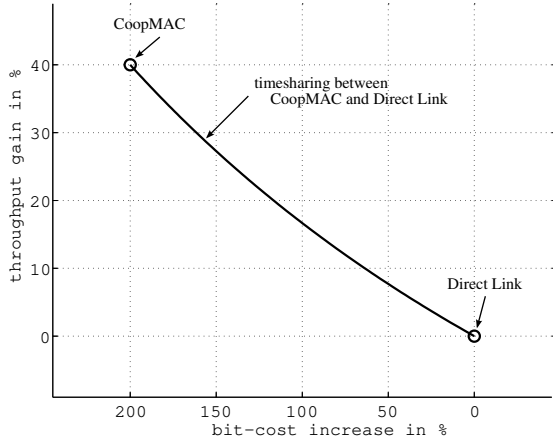}
\caption{The throughput/bit-cost tradeoff for the potential helper $n_3$ in the network displayed in Figure~\ref{fig:toyExample}. The horizontal axis displays the bit-cost increase in percent compared to Direct Link. The vertical axis displays the throughput gain in percent compared to Direct Link. In the figure, the operating points of CoopMAC and Direct Link and the tradeoff curve that results from timesharing between these two strategies is displayed.}
\label{fig:helperTradeoff}
\end{figure}

We now consider the simple network displayed in Figure~\ref{fig:toyExample}. Three nodes $n_1$, $n_2$, and $n_3$ want to transmit to the same AP. All nodes use the transmit power of $\pwr=1$~W. The rates are
\begin{align}
R_{n_1}=R_{n_2}=1\,\frac{\text{bit}}{\text{s}},\quad R_{n_1n_3}=R_{n_2n_3}=R_{n_3}=3\,\frac{\text{bit}}{\text{s}}.
\end{align}
For simplicity, we omit units in the following. Because of $1/3+1/3<1$, according to \eqref{eq:helper}, $n_3$ is a potential helper for both $n_1$ and $n_2$. For clear exposure, we postpone distributed scheduling through random access to the following sections \ref{sec:protocol} and \ref{sec:analysis} and schedule transmissions through Round Robin. The nodes $n_1$, $n_2$, and $n_3$ transmit one at a time in the fixed order $n_1,n_2,n_3,n_1,n_2,n_3,\dotsc$. In Direct Link, each node transmits one bit at a time directly to the AP, which takes the travel time $1$ for nodes $n_1$ and $n_2$ and the travel time $1/3$ for node $n_3$. In CoopMAC, nodes $n_1$ and $n_2$ first transmit their bits to $n_3$, which takes the time $1/3$. After receiving a bit from $n_1$ or $n_2$, node $n_3$ immediately forwards the received bit to the AP, which again takes the time $1/3$. Thus, the travel time in CoopMAC for bits of $n_1$ and $n_2$ is $1/3+1/3=2/3$ and for $n_3$, it is $1/3$. We can now use \eqref{eq:RRperformance} to calculate throughput and bit-cost of Round Robin based Direct Link and CoopMAC. For Direct Link, we get
\begin{align}
\thrp^\mathrm{dir} = \frac{1}{\frac{1}{1}+\frac{1}{1}+\frac{1}{3}}=\frac{3}{7},\quad\btcs^\mathrm{dir}_{n_1} = \btcs^\mathrm{dir}_{n_2} = 1,\quad \btcs^\mathrm{dir}_{n_3} = \frac{1}{3}.
\end{align}
For CoopMAC, we get
\begin{align}
\thrp^\mathrm{coop} = \frac{1}{\frac{2}{3}+\frac{2}{3}+\frac{1}{3}}=\frac{3}{5},\quad \btcs^\mathrm{coop}_{n_1} = \btcs^\mathrm{coop}_{n_2} = \frac{1}{3},\quad \btcs^\mathrm{coop}_{n_3} = 1.
\end{align}
As we can see, cooperation increases throughput from $3/7$ to $3/5$ and decreases the \emph{average} bit-cost from $7/9$ to $5/9$. However, the bit-cost of the helping node $n_3$ increases because of cooperation from $1/3$ to $1$: 
\emph{From the perspective of the helping node $n_3$, there is a tradeoff between throughput and bit-cost.}
Through timesharing between CoopMAC and Direct Link, any other operating point in-between can be made available to $n_3$. We use CoopMAC for the fraction of time $\alpha$ and Direct Link for the fraction of time $1-\alpha$. The average power of $n_3$ in CoopMAC and Direct Link is respectively
\begin{align}
\bar{\pwr}^\mathrm{coop}_{n_3}=\frac{3}{5}\quad\text{and}\quad\bar{\pwr}^\mathrm{dir}_{n_3}=\frac{1}{7} 
\end{align}
and following \eqref{eq:RRperformance}, we get for $n_3$ the tradeoff-curve parameterized by $\alpha$
\begin{align}
\thrp^\alpha = \alpha \thrp^\mathrm{coop} + (1-\alpha) \thrp^\mathrm{dir},\quad \btcs^\alpha_{n_3} = \frac{\alpha \bar{\pwr}^\mathrm{coop}_{n_3}+(1-\alpha)\bar{\pwr}^\mathrm{dir}_{n_3}}{\alpha \thrp^\mathrm{coop} + (1-\alpha) \thrp^\mathrm{dir}}\label{eq:tradeoffCurve}
\end{align}
A plot can be found in Figure~\ref{fig:helperTradeoff}. The timesharing parameter $\alpha$ determines the degree of cooperation in the network: for $\alpha=1$, the potential helpers are fully cooperative (CoopMAC) and for $\alpha=0$, the potential helpers do not cooperate at all (Direct Link).

\section{CSMA Based fairMAC Protocol}
\label{sec:protocol}
For sake of clarity, we make some simplifying assumptions
for the MAC layer. Since we are interested in high throughput for all nodes, we assume that all nodes
operate in saturation mode, i.e., they are backlogged and we do not need to
consider packet arrival processes in our analysis. We consider
slotted CSMA with the two parameters slot length $\sigma$ and transmit
probability $\tau$. In wireless networks, there are several reasons for packet losses. We include in
our work packet losses because of interference (collision) but neglect
other forms of packet losses. By assigning appropriate probabilities to other kinds of packet losses, they can be incorporated into our model, e.g., packet losses because of deep fade are considered in an ongoing research project at our institute. We further assume that control headers and acknowledgments (ACK) are transmitted at a base rate and that they can be decoded by all nodes in the 
network. To remain general, we assume that data packets are large enough such
that the specific size of control data is negligible. Finally, we assume
that ACKs never get lost.
\subsection{Reference Protocols}
We start by defining the two reference protocols Direct Link and CoopMAC. While we used both terms for Round Robin based strategies in the previous section, they refer to CSMA based strategies here and hereafter, if not explicitly stated differently.
\subsubsection{Direct Link \cite{Bianchi2000}}
\label{subsec:directlink}
When node $k$ seeks to transmit a packet, it competes for the medium according
to CSMA:~if $k$ senses the channel idle in time slot $m$, it initiates a
transmission with probability $\tau$ in time slot $m+1$. If no other node is
transmitting at the same time, the AP can decode the packet and sends an ACK in return. Otherwise, a collision occurs; no ACK is sent by the AP; node $k$ declares its packet lost and will try to transmit again the same packet later.
\subsubsection{CoopMAC in base mode \cite{Liu2007}}
\label{subsec:coopmac}
All nodes initiate the transmission of an own packet in the same way as in
Direct Link. Assume that node $k$ initiates a transmission. We have to
distinguish two situations.\\
$\bullet$ Node $k$ has no helper. The transmission is performed according to the
Direct Link protocol.\\
$\bullet$ Node $k$ has a helper $h$. In this case, $k$ transmits its packet to $h$
at rate $R_{kh}$. If $h$ can decode the packet, it immediately forwards the
packet to the AP at rate $R_{h}$. The AP sends an ACK to $k$. If $h$ cannot
decode the packet because of collision, it remains idle. Node $k$ detects the collision by not receiving the ACK. Node $k$ declares its packet lost and
tries to transmit the same packet again via $h$ later.
\subsection{fairMAC}
\label{subsec:fairmac}
CoopMAC was designed to maximize throughput. However, the resulting
bit-cost of potential helping nodes compared to other nodes can become very large, as we have seen in the previous section. Although a node addressed for help can in principal refuse to help, bit-cost control at helping nodes is not incorporated in CoopMAC. This is because the source $k$ decides when the helper $h$ has to help: $h$ forwards immediately the packet from $k$. In fairMAC, this decision is taken by $h$: node $h$ stores the data from $k$ and
transmits it in conjunction with one of his own future packets. This procedure
is managed in a distributed manner at source $k$ and helper $h$ as follows.\\
$\bullet$ \textbf{Helping node} $h$ manages an additional, infinite packet queue for the packets to be forwarded. When $h$ receives a packet from $k$, $h$ adds it to this queue and notifies $k$ by sending a ``preACK'' to $k$. When node $h$ initiates a transmission
to the AP, it forms a joint packet consisting of own data from its buffer and data of up
to $Q$ packets from the forwarding queue. If there is no collision, the AP
successfully decodes the joint packet and sends one ``jointACK'' to $h$ and all other nodes with data in the joint packet. Node $h$ receives
the jointACK and removes the corresponding packets from the forwarding queue.\\
$\bullet$ \textbf{Source node} $k$ tracks the packet delay at helper $h$ by a state
variable $p$ that indicates the number of pending packets. Each time $k$
transmits a packet to $h$ and receives a preACK, it increases $p$ by one.
When $p$ passes the maximum number of pending packets $P$, $k$ directly transmits its current packet to the AP. When $k$ receives a jointACK from the AP, it decreases $p$ by the number of its pending packets that helper $h$ finally forwarded to the AP in the corresponding joint packet.

\section{Theoretical Analysis}
\label{sec:analysis}
In this section, we derive analytic formulas for the throughput and bit-cost of Direct Link and CoopMAC and show how these are related to the corresponding values of fairMAC. Our derivations are inspired by \cite{Bianchi2000}, but we follow the notation in \cite{Bocherer2008}.
\subsection{Throughput and Bit-Cost of Direct Link and CoopMAC}

The network situation over time can be split into phases. In each phase, the network can either be idle, there can be a successful transmission, or there can be a collision.
In the average, one network phase is idle for the time $\bar{t}_i$, it consists in a successful transmission for the time of $\bar{t}_s$, and it consists in a collision for the time of $\bar{t}_c$. We normalized the amount of data of one successful transmission to $1$~bit. As a result, both packet durations and slot time $\sigma$ have to be normalized by the number of bits in a typical packet. For now, this observation is not of further importance, however, we will come back to this observation in Subsection~\ref{subsec:throughputBitCostFairMAC}.
The probability $p_s$ that one specific node $k$ transmits 
successfully in a given phase is given by
\begin{align}
p_s = \tau(1-\tau)^{N-1}.
\end{align}
Therefore, the throughput $\thrp$ per node is given by
\begin{align}
\thrp&=\frac{p_s}{\bar{t}_s+\bar{t}_c+\bar{t}_i}\label{eq:throughput}.
\end{align}
We now explicitly calculate $\bar{t}_s$, $\bar{t}_i$, $\bar{t}_c$ in \eqref{eq:throughput}. The
probability of an idle phase is
\begin{align}
p_i = (1-\tau)^N.
\end{align}
The time $\bar{t}_i$ a phase is idle in the average is given by
\begin{align}
\bar{t}_i = p_i\sigma.\label{eq:ti}
\end{align}
The travel time as introduced in Subsection~\ref{subsec:throughputBitCost} is the duration one packet needs to travel from the source node to the destination.
For node $k$, it is given by $s_k=1/R_{kh_k}+1/R_{h_k}$ if $k$ has helper $h$ and it is given by $s_k=1/R_k$ if $k$ transmits directly to the AP. 
The average time of successful transmission in one phase is now given by
\begin{align}
\bar{t}_s = \sum_{k=1}^N p_s(s_k+\sigma).\label{eq:ts}
\end{align}
Here, we have to add the slot length $\sigma$ to the travel time $s_k$ since every transmission is followed by an idle slot (no node will transmit right after an ongoing transmission in CSMA since it first needs to sense an idle slot). It remains to calculate the average collision time $\bar{t}_c$. In CoopMAC, we do not need to consider forwarding transmissions from helpers to the AP, because helping nodes only forward packets if there was no collision in the first hop. Since forwarding happens immediately, there cannot be a collision in the second hop, see Subsection~\ref{subsec:coopmac}). Relevant for the collision time is thus the \emph{packet duration}, which we denote by $u_k$. If node $k$ transmits via $h_k$, $u_k=1/R_{kh_k}$ and if $k$ transmits directly to the AP, $u_k=1/R_k$.
We assume without loss of generality that the set of packet lengths $\{u_k\}_{k=1,\dotsc,N}$ is ordered, i.e., $k<l\Rightarrow u_k\leq u_l$. The average collision time $\bar{t}_c$ is then given by
\begin{align}
\bar{t}_c &= \sum\limits_{k=2}^N
\underbrace{\tau(1-\tau)^{N-k}}_{\text{(i)}}\sum\limits_{l=1}^{k-1}
\underbrace{\binom{k-1}{l}\tau^l(1-\tau)^{k-1-l}}_{\text{(ii)}}
(u_k+\sigma)
\label{eq:tc}
\end{align}
where the term (i) is the probability that node $k$ transmits and nodes with packet length larger than $u_k$ (and possibly some nodes with packet length equal to $u_k$) do not transmit, and where the term (ii) is the probability that exactly $l$ nodes with packet length smaller than or equal to $u_k$ transmit. Using \eqref{eq:ti}, \eqref{eq:ts}, and \eqref{eq:tc} in \eqref{eq:throughput} allows us to explicitly calculate the average per node throughput $\thrp$ of CSMA based CoopMAC and Direct Link for a given network.
%%%% Bit-Cost

To calculate the bit-cost of node $k$, we need to differ between two kinds of transmission: first, transmitting own data to the AP (or the helper), and second, forwarding data of other nodes to the AP. As stated above, transmission of own data is involved in collisions and forwarding is not. Therefore, to successfully transmit one own bit, node $k$ has to try $\tau/p_s$ times and to successfully forward a received packet, node $k$ only needs to try once. As introduced in Subsection~\ref{subsec:throughputBitCost}, $H_k$ denotes the number of nodes that get help from node $k$. In the average, node $k$ forwards $H_k$ packets per own successfully transmitted bit and the resulting bit-cost of node $k$ is
\begin{align}
\btcs_k&=\bigl(H_k+\frac{\tau}{p_s}\bigr) u_k \pwr\label{eq:cost}
\end{align}
where $\pwr$ is the transmission power, according to our system model from Section~\ref{sec:model}. Note that in Direct Link, $H_k=0$ for all nodes $k=1,\dotsc,N$ in the network. 
%%%%%%%%%%%%%%%%%%%%%%
\subsection{Throughput and Bit-Cost of fairMAC}
\label{subsec:throughputBitCostFairMAC}
%%%%%%%%%%%%%%%%%%%%%%
We now relate two specific configurations of fairMAC to CoopMAC and Direct Link. First, when the maximum number of pending packets $P$ is finite and  the maximum number $Q$ of packets forwarded at a time is equal to zero (no cooperation at all), which we refer to by fairMAC$_0$. Second, when $P=Q=\infty$, which we refer to by fairMAC$_\infty$. The values $P=Q=\infty$ may appear unrealistic, however, as we will see in Section~\ref{sec:simulation}, the theoretical behavior of fairMAC$_\infty$ can already be observed for moderate values of $P$ and $Q$, which makes the investigation of fairMAC$_\infty$ reasonable. We denote the throughput and bit-cost of CoopMAC and Direct Link as can be calculated by \eqref{eq:throughput} and \eqref{eq:cost} in the following by $\thrp^\mathrm{coop}$, $\btcs^\mathrm{coop}_k$, $\thrp^\mathrm{direct}$, $\btcs^\mathrm{direct}_k$,  respectively. The corresponding values for fairMAC$_0$ and fairMAC$_\infty$ are denoted by $\thrp^{\mathrm{fair}_0}$, $\btcs^{\mathrm{fair}_0}_k$ and $\thrp^{\mathrm{fair}_\infty}$, $\btcs^{\mathrm{fair}_\infty}_k$.
\begin{proposition}
\label{prop:qzero}
fairMAC$_0$ reaches the Direct Link operating point, i.e.,
\begin{align}
\thrp^{\mathrm{fair}_0} = \thrp^\mathrm{direct}\quad\text{and}\quad \btcs^{\mathrm{fair}_0}_k = \btcs^\mathrm{direct}_k.
\end{align}
\end{proposition}
\begin{IEEEproof}
In fairMAC$_0$, nodes that try to transmit via their helper loose their first $P$ packets, since these packets are transmitted to the corresponding helpers but then never forwarded because of $Q=0$. After that, the number of pending packets is $p=P$ and all nodes will transmit all following packets directly to the AP, which happens exactly according to Direct Link (this can be seen from the protocol descriptions in Section~\ref{sec:protocol} by setting $p=P$ and $Q=0$). On the long term, the impact of the lost $P$ packets onto throughput and bit-cost gets infinitesimal small and the proposition follows.
\end{IEEEproof}
We now prepare for the investigation of fairMAC$_\infty$. The two parameters transmission probability $\tau$ and slot length $\sigma$ are network parameters, which take different values depending on which network setup we consider. We assign to the transmit probability a value $\tau\propto\sqrt{\sigma}$ and let then $\sigma$ got to zero. Note that it was shown in \cite{Bianchi2000} that $\tau\propto\sqrt{\sigma}$ also holds for that $\tau$ that maximizes throughput for a given $\sigma$; our assignment is thus reasonable. Since $\sigma$ is normalized by the number of bits in a packet, letting $\sigma$ go to zero in our formulas corresponds to letting the packet duration go to infinity in the corresponding real world system.
\begin{proposition}
\label{prop:sigmaZero}
For $\tau\propto\sqrt{\sigma}$ and $\sigma\rightarrow 0$, CSMA based CoopMAC and Direct Link perform asymptotically as Round Robin based CoopMAC and Direct Link (see \eqref{eq:RRperformance}), i.e.,
\begin{align}
\thrp^* &= \lim_{\sigma\rightarrow 0} \thrp = \frac{1}{\sum_{k=1}^N s_k}\label{eq:throughputSigmaZero}\\
\btcs^*_k &= \lim_{\sigma\rightarrow 0} \btcs_k = (H_k+1)u_k\pwr
\end{align}
\end{proposition}
\begin{IEEEproof}
We write
\begin{align}
\thrp=\frac{p_s}{\bar{t}_s+\bar{t}_c+\bar{t}_i}
&=\frac{\tau(1-\tau)^{N-1}}{\bar{t}_s+\bar{t}_c+\bar{t}_i}\\
&=\frac{(1-\tau)^{N-1}}{\frac{\bar{t}_s}{\tau}+\frac{\bar{t}_c}{\tau}+\frac{\bar{t}_i}{\tau}}\label{eq:propProof}
\end{align}
As $\tau\propto\sqrt{\sigma}$, $\tau\overset{\sigma\rightarrow 0}{\longrightarrow} 0$ and $\frac{\sigma}{\tau}\overset{\sigma\rightarrow 0}{\longrightarrow} 0$. Using these two limits, it follows through some basic arithmetic operations that for $\sigma\rightarrow 0$, the numerator of the right-hand side of \eqref{eq:propProof} converges to $1$ and that the denominator of \eqref{eq:propProof} converges to $\frac{1}{\sum_{k=1}^N s_k}$. Thus \eqref{eq:throughputSigmaZero} follows. For the bit-cost, we have from \eqref{eq:cost}
\begin{align}
\btcs_k=\bigl(H_k+\frac{\tau}{p_s}\bigr) u_k \pwr
&=\bigl(H_k+\frac{\tau}{\tau(1-\tau)^{N-1}}\bigr) u_k\pwr\\
&=\bigl(H_k+\frac{1}{(1-\tau)^{N-1}}\bigr) u_k\pwr.
\end{align}
The right-hand side of the last line converges to $(H_k+1)u_k\pwr$ for $\tau\rightarrow 0$. This concludes the proof.
\end{IEEEproof}

To derive throughput and bit-cost expressions for fairMAC$_\infty$, we first calculate the packet lengths $v_k$ in fairMAC$_\infty$. Denote by $\mathcal{D}$ the set of nodes that transmit their packets directly to the AP, denote by $\mathcal{H}\subseteq \mathcal{D}$ the set of nodes that help at least one other node, and denote by $\mathcal{C}$ the set of nodes that transmit their packets via a helper. In fairMAC$_\infty$, the packet lengths $v_k$ of nodes $k\in\mathcal{D}\setminus\mathcal{H}$ are deterministic values given by $v_k=1/R_k$. Since $P=\infty$, nodes $k\in\mathcal{C}$ will always transmit via their helper, and the packet lengths are also deterministic values given by $v_k=1/R_{kh_k}$. The packet length $v_k$ for nodes $k\in\mathcal{H}$ are random values given by $(X_k+1)/R_k$ where $X_k$ is the number of packets in the forwarding queue of node $k$ right before $k$ is transmitting. We can now see that deriving the expression for the throughput $\thrp^{\mathrm{fair}_\infty}$ when $\sigma$ is non-zero is intricate, because packets involved in collisions are of varying length. We defer these calculations to an extended version of this work. However, it can easily be seen that the term corresponding to the right-hand side of \eqref{eq:throughputSigmaZero} is for fairMAC$_\infty$ given by
\begin{align}
\thrp^{*\mathrm{fair}_\infty} = \frac{1}{\expop[\sum_{k=1}^\infty v_k]}\label{eq:throughputSigmaZeroFair}
\end{align}
Since the random variables $\{X_k\}_{k\in\mathcal{H}}$ are mutually independent, we can exchange summation and expectation in \eqref{eq:throughputSigmaZeroFair}. The expectation of $X_k$ is $\expop[X_k]=H_k$. We get
\begin{align}
&\thrp^{*\mathrm{fair}_\infty} = \frac{1}{\expop\left[\sum_{k=1}^N v_k\right]}\\
&= \frac{1}{\sum_{k \in \mathcal{D}\setminus\mathcal{H}}\frac{1}{R_k}+\sum_{k\in\mathcal{C}}\frac{1}{R_{kh_k}}+\sum_{k\in\mathcal{H}} \expop[v_k]}\\
&= \frac{1}{\sum_{k\in \mathcal{D}\setminus\mathcal{H}}\frac{1}{R_k}+\sum_{k\in\mathcal{C}}\frac{1}{R_{kh_k}}+\sum_{k\in\mathcal{H}} (1+H_k)\frac{1}{R_k}}.
\end{align}
We can now reorder the terms and get, continuing from the last line
\begin{align}
&\frac{1}{\sum_{k\in \mathcal{D}\setminus\mathcal{H}}\frac{1}{R_k}+\sum_{k\in\mathcal{C}}\frac{1}{R_{kh_k}}+\sum_{k\in\mathcal{H}} (1+H_k)\frac{1}{R_k}}\\
&= \frac{1}{\sum_{k\in \mathcal{D}}\frac{1}{R_k}+\sum_{k\in\mathcal{C}}\frac{1}{R_{kh_k}}+\sum_{k\in\mathcal{H}} H_k\frac{1}{R_k}}\\
&= \frac{1}{\sum_{k\in \mathcal{D}}\frac{1}{R_k}+\sum_{k\in\mathcal{C}}(\frac{1}{R_{kh_k}}+\frac{1}{R_{h_k}})}\\
&= \thrp^{*\mathrm{coop}}.\label{eq:resultThroughputInfty}
\end{align}
Since $\expop[X_k]=H_k$, we can directly express the bit-cost of fairMAC$_\infty$ for $\sigma=0$ by the packet lengths $u_k$ in CoopMAC as
\begin{align}
\btcs^{*\mathrm{fair}_\infty}_k &= \expop[v_k]\pwr
=
\left\lbrace
\begin{array}{ll}
1/R_k\pwr=u_k\pwr,&k\in\mathcal{D}\\
1/R_{kh_k}\pwr=u_k\pwr,&k\in\mathcal{C}\\
\frac{H_k+1}{R_k}\pwr=(H_k+1)u_k\pwr,&k\in\mathcal{H}
\end{array}
\right.
\end{align}
\begin{align}
&=(H_k+1)u_k\pwr\label{eq:bitCostInfty}\\
&=\btcs^{*\mathrm{coop}}_k.\label{eq:resultBitCostInfty}
\end{align}
To see that equality in \eqref{eq:bitCostInfty} holds, remember that $H_k=0$ if node $k$ does not help any other node. Combining result \eqref{eq:resultThroughputInfty} and result \eqref{eq:resultBitCostInfty} with Proposition~\ref{prop:sigmaZero}, we have shown
\begin{proposition}
\label{prop:sigmaZeroFairMAC}
fairMAC$_\infty$ asymptotically reaches the operating point of Round Robin based CoopMAC when $\sigma\rightarrow 0$, i.e.,
\begin{align}
\thrp^{*\mathrm{fair}_\infty} = \thrp^\mathrm{*coop}\quad\text{and}\quad \btcs^{*\mathrm{fair}_\infty}_k = \btcs^{*\mathrm{coop}}_k.
\end{align}
\end{proposition}

\section{Numerical Results and Discussion}
\label{sec:simulation}
We validate our results from Section~\ref{sec:analysis} by simulation via an implementation of fairMAC as defined in \ref{sec:protocol} in our custom network simulator written in object-oriented MATLAB. We compare the empirical values for throughput and bit-cost of fairMAC to the theoretical values of CoopMAC and Direct Link as obtained from \eqref{eq:throughput} and \eqref{eq:cost}. The tradeoff curve between these two is calculated by \eqref{eq:tradeoffCurve}. We call it in the following the \emph{timesharing curve}.

\begin{figure}
\centering
\footnotesize
\psfrag{Q = 0}{$Q=0$}
\psfrag{Q = 1}{$Q=1$}
\psfrag{Q = 2}{$Q=2$}
\psfrag{Q = 4}{$Q=4$}
\psfrag{s = 0.0088}{$\sigma=0.0088$}
\psfrag{s = 0.0001}{$\sigma=0.0001$}
\psfrag{t = 0.045}{$\tau=0.045$}
\psfrag{t = 0.0033}{$\tau=0.0033$}
\includegraphics[width=0.48\textwidth]{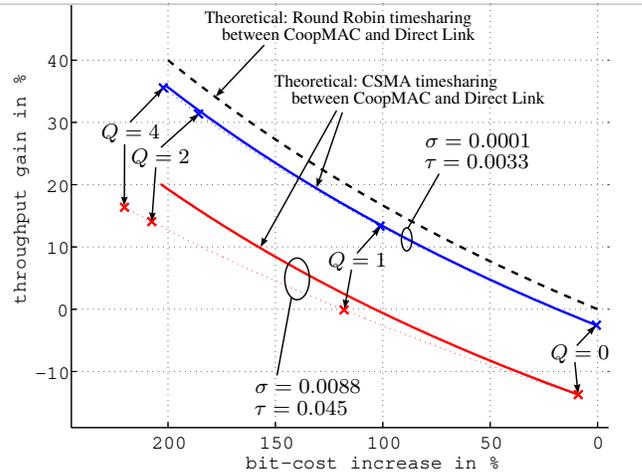}
\caption{The throughput/bit cost tradeoff of node $n_3$ in the network from Figure~\ref{fig:toyExample}. The reference strategy is for all displayed values the theoretical value for Round Robin Direct Link. For a discussion of the plot, see Section~\ref{sec:simulation}.}
\label{fig:toyCSMA}
\end{figure}

We simulate fairMAC for the network from Figure~\ref{fig:toyExample} and consider the throughput/bit-cost tradeoff at the potential helper $n_3$. We let the nodes compete $30\,000$ times for the channel. We set the maximum number $P$ of pending packets constantly equal to $P=10$ in all simulations. For the number $Q$ of packets forwarded at a time by the helping node $n_3$, we evaluate fairMAC for the values $Q=0,1,2,4$. As can be seen in Figure~\ref{fig:toyCSMA}, fairMAC reaches for $Q=0$ the corresponding CSMA Direct Link operating point, both for the network parameters $(\sigma,\tau)=(0.0088,0.045)$ and $(\sigma,\tau)=(0.0001,0.0033)$. This validates Proposition~\ref{prop:qzero}. For the typical value $\sigma=0.0088$ (see \cite{Bocherer2008}), the CSMA timesharing curve is far away from the Round Robin timesharing curve. However, for the smaller value $\sigma=0.0001$, the CSMA curve is close to the Round Robin curve. This validates Proposition~\ref{prop:sigmaZero}. For $\sigma=0.0088$, the operating points for $Q=2$ and $Q=4$ of fairMAC are below the corresponding CSMA time sharing curve. This is because in fairMAC, the helping nodes eventually transmit long packets (e.g, for $Q=4$ helping nodes can forward up to $4$ packets at a time). These long packets can be involved in collisions, which is expensive both in terms of throughput and bit-cost. For the smaller value $\sigma=0.0001$, all fairMAC operating points get close both to the corresponding CSMA timesharing curve and the Round Robin timesharing curve. This validates Proposition~\ref{prop:sigmaZeroFairMAC}. We conclude that our distributed protocol fairMAC can asymptotically reach the Round Robin timesharing curve as $\sigma$ goes to zero, which corresponds in real world to packet lengths going to infinity.

\section{Conclusions}
Under the throughput fairness constraint, we identified a tradeoff between throughput and bit-cost in CSMA based cooperative networks. From the helper node perspective, no cooperation (Direct Link) is optimum in terms of bit-cost while always cooperating (CoopMAC) is optimum in terms of throughput. We proposed the new distributed cooperative CSMA protocol fairMAC and showed both theoretically and by simulation that fairMAC can reach the throughput/bit-cost tradeoff curve.

\bibliographystyle{IEEEtran}

\bibliography{IEEEabrv,confs-jrnls,Literatur}

\end{document}